\documentclass[twocolumn,twoside,slac_two]{revtex4}
\usepackage{graphicx}
\usepackage{fancyhdr}
\usepackage{amsmath}
\begin{document}

\title{Gamma-ray lines in the Fermi-LAT data?}

\author{C. Weniger}
\affiliation{GRAPPA Institute, Univ.~of Amsterdam, Science Park 904,
1098 GL Amsterdam, Netherlands}

\begin{abstract}
  Gamma-ray lines are traditional smoking gun signatures for dark matter
  annihilation in the Universe. In regions optimized for large signal-to-noise
  ratio, we identified a signal candidate for a gamma-ray line at energies
  around 130 GeV with a post-trials significance of $3.2\sigma$.  Spectral and
  spatial properties are not inconsistent with a dark matter signal. One year
  has passed since the initial papers, and I give here a brief summary and an
  update on the status of the 130 GeV feature in the un-reprocessed P7
  gamma-ray data of the Fermi-LAT.
\end{abstract}

\maketitle

\thispagestyle{fancy}

\section{Introduction}
One of the main challenges in searches for a signal from dark matter
annihilation is the signal-background discrimination. The dark matter signal is
expected to be faint in most scenarios, and often much weaker than the
systematic uncertainties associated with the Galactic diffuse emission and
unresolved point sources. Smoking-gun signatures like gamma-ray lines, or
related sharp features from internal Bremsstrahlung processes or cascade
decays, could hence become the cornerstone for an unequivocal discovery of a
dark matter signal in the gamma-ray sky.

Our recent identification of a line-like feature with a post-trials
significance of $3.2\sigma$ around energies of 130 GeV in the Galactic center
data of the Fermi Large Area Telescope (LAT) (\cite{Bringmann:2012vr} and
\cite{Weniger:2012tx}) gave rise to a large number of papers, studying possible
explanations in terms of dark matter annihilation, mono-energetic pulsar winds
(\cite{Aharonian:2012cs}), and instrumental effects (\cite{Finkbeiner:2012ez,
Hektor:2012ev, Whiteson:2012hr, Whiteson:2013cs}).  The signal candidate was
subsequently confirmed independently by \cite{Tempel:2012ey} and
\cite{Su:2012ft} (for a recent review on indirect dark matter searches with
gamma rays see \cite{Bringmann:2012ez}).

In this proceedings contribution, I will briefly summarize the results from the
initial analysis and some of the follow-up work in section~\ref{sec:init}, give
an update in section~\ref{sec:update}, and conclude in
section~\ref{sec:conclusions}.

\section{Identification and developments}
\label{sec:init}
Most searches for gamma-ray lines from WIMP (weakly interacting massive
particle) annihilation are technically a shape analysis of the gamma-ray flux
measured in regions of interest (ROIs) with a large signal-to-noise ratio for
dark matter signals. The spectral analysis is usually confined to a small
energy range around the line energy of interest, such that the smooth
background spectra can then be approximated by a single power-law. The basic
idea is here to trade systematic uncertainties in the background flux for
statistical errors.

\begin{figure*}
  \includegraphics[width=.42\textwidth]{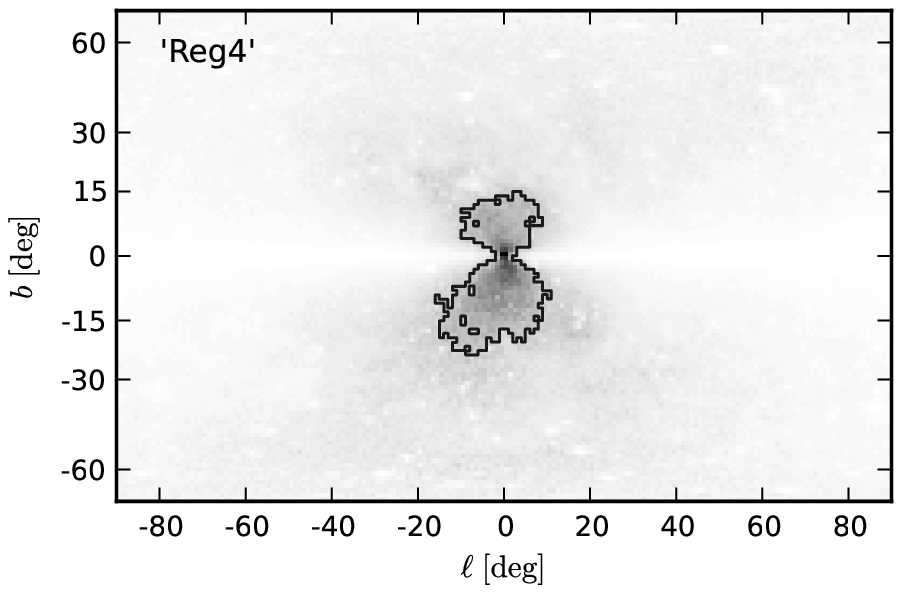}
  \includegraphics[width=.45\textwidth]{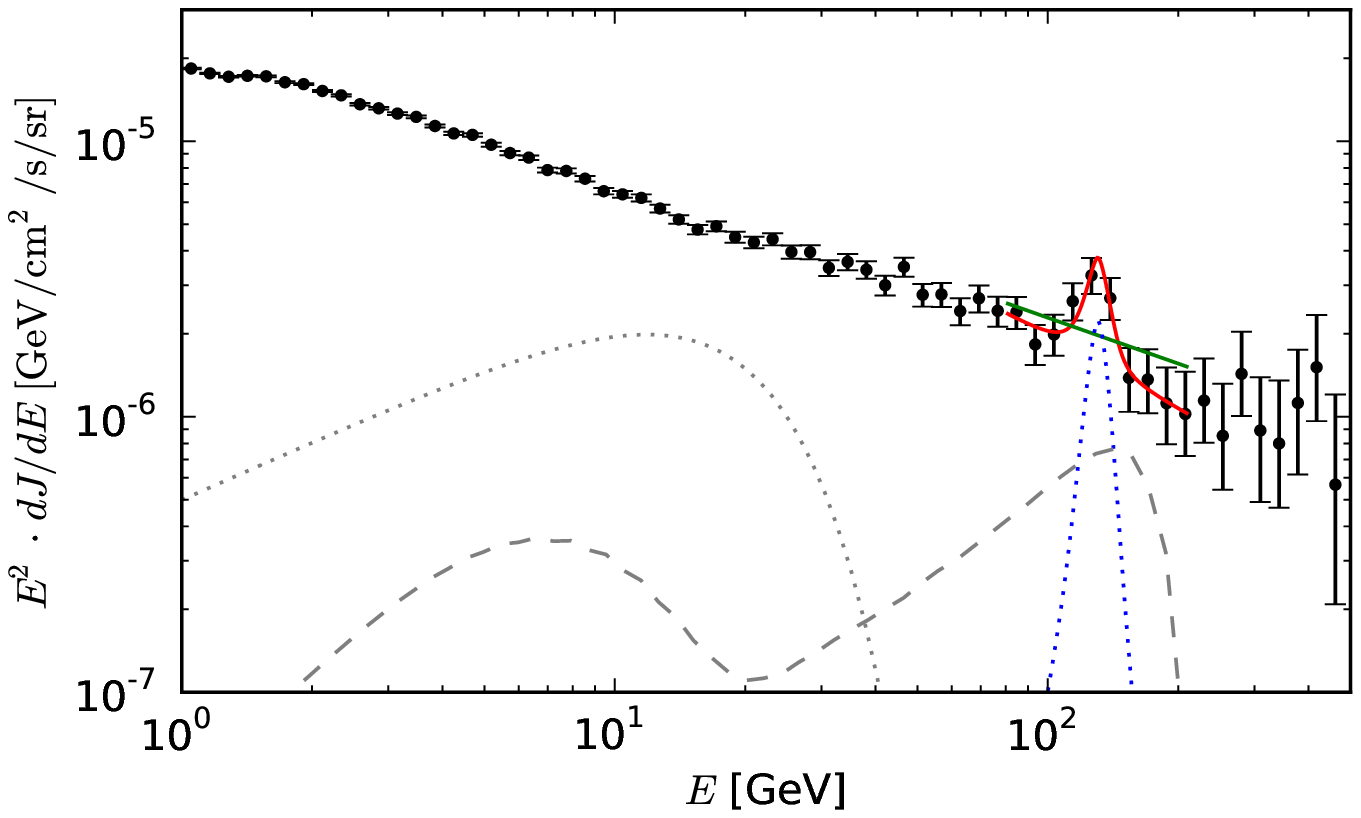}
  \caption{\emph{Left panel:} ROI Reg4 from \cite{Weniger:2012tx}, optimized
  for large S/N in case of slightly contracted profile. \emph{Right panel:}
  Gamma-ray flux measured within that ROI by Fermi-LAT. An excess of events
  around 130 GeV is clearly visible in the data. We show the fits to the data
  in the energy range 80--210 GeV. For direct comparison we show a very hard
  spectrum with super-exponential cut-off (left dotted line; $\sim
  E^{-1.3}\exp[-(E/20{\ \rm GeV})^2]$) and the ICS emission from mono-energetic
  230 GeV electrons at the Galactic center (dashed), both with arbitrary
  normalization.}
  \label{fig:regSpec}
\end{figure*}

In \cite{Bringmann:2012vr, Weniger:2012tx}, we used an adaptive method to find
ROIs optimized for different profiles of the Galactic dark matter halo. As
template for the background morphology, we took the gamma-ray flux measured at
1--20~GeV; for a given dark matter signal profile, the optimal ROI was than
uniquely determined using a simple deterministic algorithm that optimizes the
signal-to-noise ratio.  For non-cored dark matter profiles, these ROIs take
roughly the shape of an hourglass. One of these ROIs (for a slightly contracted
profile) is shown in the left panel of Fig.~\ref{fig:regSpec} by the black line
(c.p.~\cite{Weniger:2012tx}).

The differential flux of gamma rays measured in this ROI is shown in the right
panel of Fig.~\ref{fig:regSpec}. Already by eye, one can identify a
surprisingly clear line-like excess at energies around $130$~GeV. The two solid
lines to the right represent a power-law only (power-law + line signal) fit to
the data, restricted to the energy range 80--210 GeV. The formal significance
for a line feature was found to be $4.6\sigma$ (\cite{Weniger:2012tx}; even
higher $>5\sigma$ significances were found in the template analysis
by~\cite{Su:2012ft}). To illustrate the sharpness of the feature, the gray
dotted line at the left shows a spectrum with super-exponential cutoff; the
gray dashed line shows the inverse Compton scattering (ICS) radiation generated
by a mono-energetic electron population, scattering with star light at the
Galactic center.  Even this highly idealized ICS emission is disfavoured
w.r.t.~to a monochromatic gamma-ray line by about $3\sigma$ (with $TS\approx12$
instead of $TS\approx21$ for the line).

If the 130 GeV signature is interpreted as a gamma-ray line produced by dark
matter annihilation via $\chi\chi\to\gamma\gamma$, the corresponding line from
$\chi\chi\to\gamma Z^0$ would be expected at a gamma-ray energy of $\simeq114$
GeV, though the strength of this line would be model dependent.  Indeed, weak
indications for such a second line at the $1\sigma$--$2\sigma$ level were found
by different groups (see e.g.~\cite{Rajaraman:2012db, Su:2012ft,
Bringmann:2012ez}).  Furthermore, it was found that the signature is extended
and roughly compatible with the flux profile expected from a conventional NFW
or Einasto dark matter profile (see \cite{Bringmann:2012ez}). At the very
center, however, the signature appears to be displaced from the Galactic center
(with $\sim2\sigma$ significance, see e.g.~\cite{Su:2012ft, Rao:2012fh}).
Whether or not this already excludes an interpretation as dark matter signal is
a subject of current debate and certainly requires more data to map the
morphology of the feature more accurately (see e.g.~\cite{Kuhlen:2012qw,
Gorbunov:2012sk}).

After identification of the signature, different groups made efforts to find
corroborating evidence for a true dark matter signal. \cite{Hektor:2012kc}
found indications for the emission of a double line (at 110 and 130 GeV
energies) when stacking 18 of the most promising galaxy clusters. The
statistical significance was estimated to be $3.6\sigma$. \cite{Su:2012zg}
identified a similar double line structure with $3.3\sigma$ significance when
stacking unassociated point sources. These point sources would then have to be
interpreted as dark matter subhalos emitting a strong annihilation signal ---
an interpretation that was challenged in \cite{Hooper:2012qc, Mirabal:2012za},
who found that the continuum spectrum of the unassociated sources in question
is not compatible with expectations from dark matter and suggested active
galactic nuclei as possible candidates. In that case, a significant line
emission from these sources would likely indicate an instrumental effect.

Earth limb photons, which stem from cosmic-ray scattering on Earth atmosphere
nuclei, provide a smooth reference spectrum for systematic checks. Most
interestingly, it was found (\cite{Finkbeiner:2012ez, Hektor:2012ev,
BloomFS:2012}) that the low-incidence angle ($\theta<60^\circ$) part of the
Earth limb data exhibits a gamma-ray line feature at 130 GeV with a
significance of about $3\sigma$. This could point towards an instrumental
effect generating 130 GeV lines. However, the same signature was not found in
other low-incidence angle test regions, like the Galactic disk (excluding the
Galactic center). Given the fact that the Galactic center signature is
localized to within a few degrees around the Galactic center, over which the
observational profile of the LAT does not change significantly, no consistent
interpretation of the Earth limb line and the Galactic center line in terms of
an instrumental effect has emerged yet. A further possible line signature at
130 GeV was found recently by \cite{Whiteson:2013cs} in a $5^\circ$ region
following the Sun.

Formally, and before trials, the most significant feature is the line at the
Galactic center. Given the completely different nature of the various targets,
it seems rather unlikely that all these signatures have a common instrumental
or physical origin (for an in-depth discussion of some of the possible
instrumental effects see \cite{Finkbeiner:2012ez}). In any case, there is no
way around waiting for additional data to see which of these signatures, if
any, are real effects and which are statistical flukes in light of a large
number of hidden trials.

\begin{figure*}[t]
  \includegraphics[width=.4\textwidth]{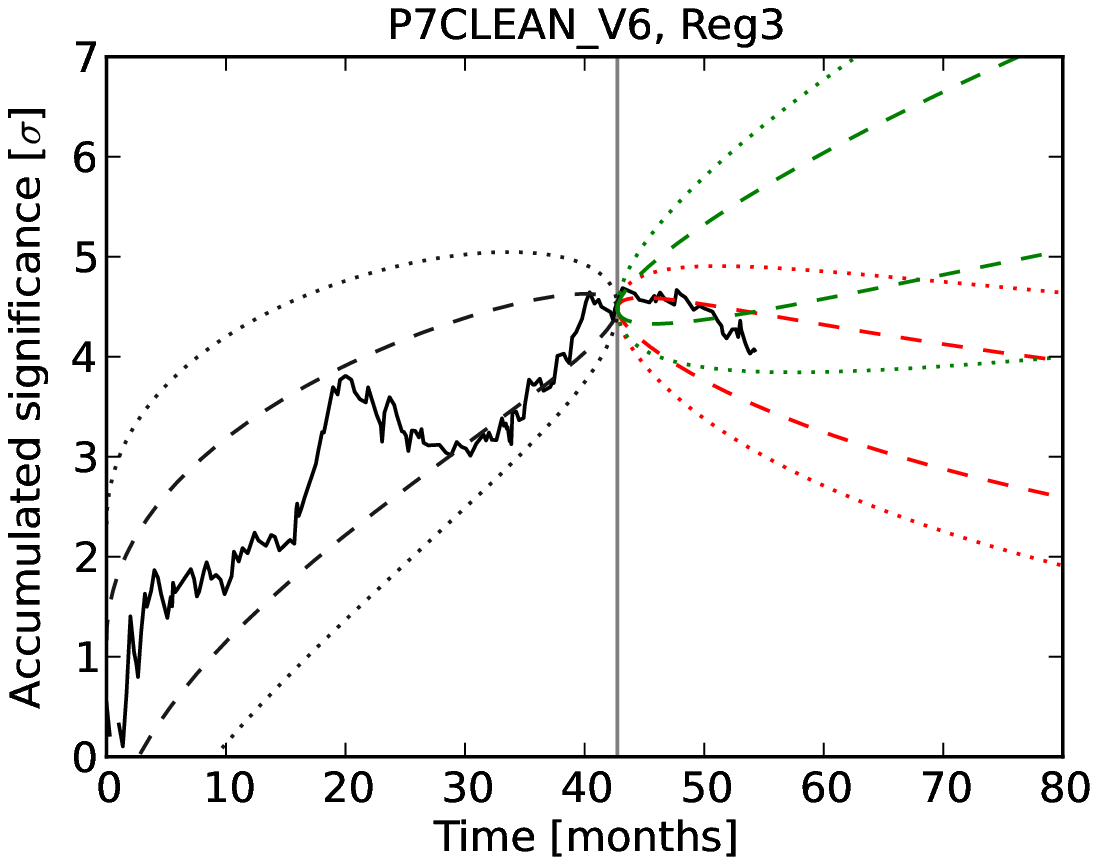}
  \includegraphics[width=.4\textwidth]{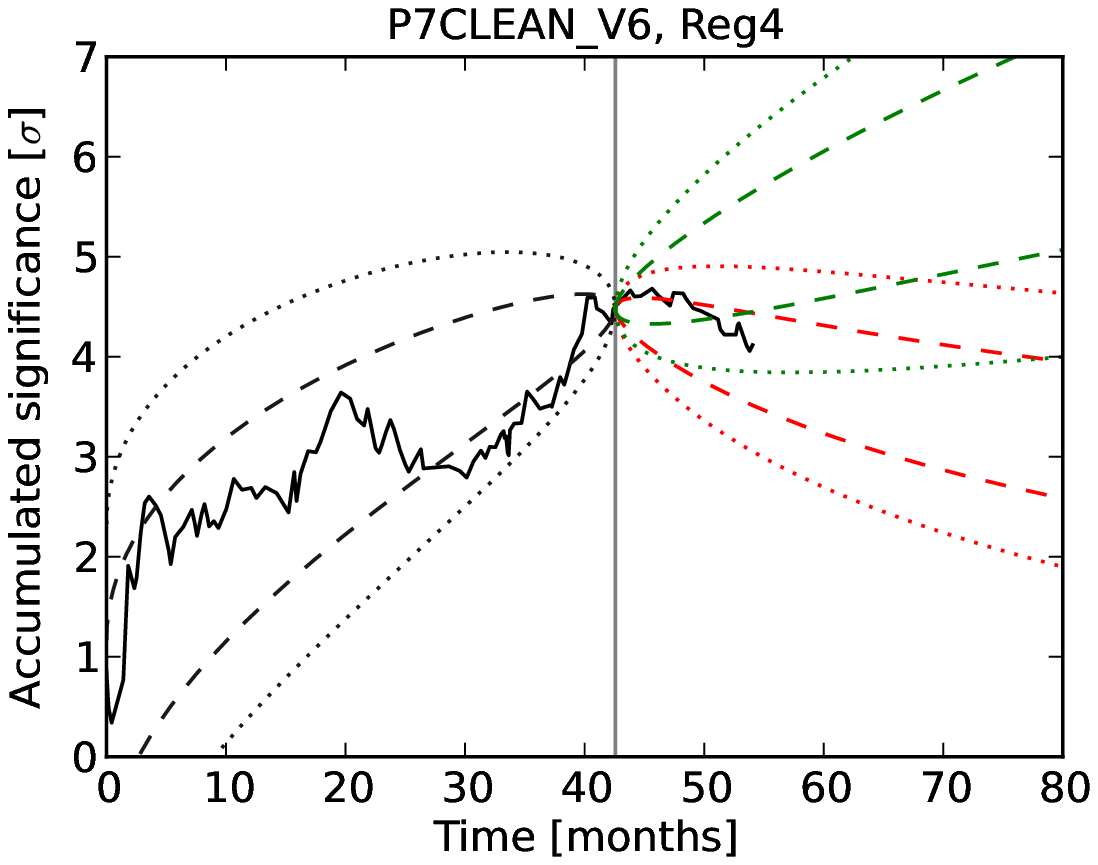}
  \includegraphics[width=.4\textwidth]{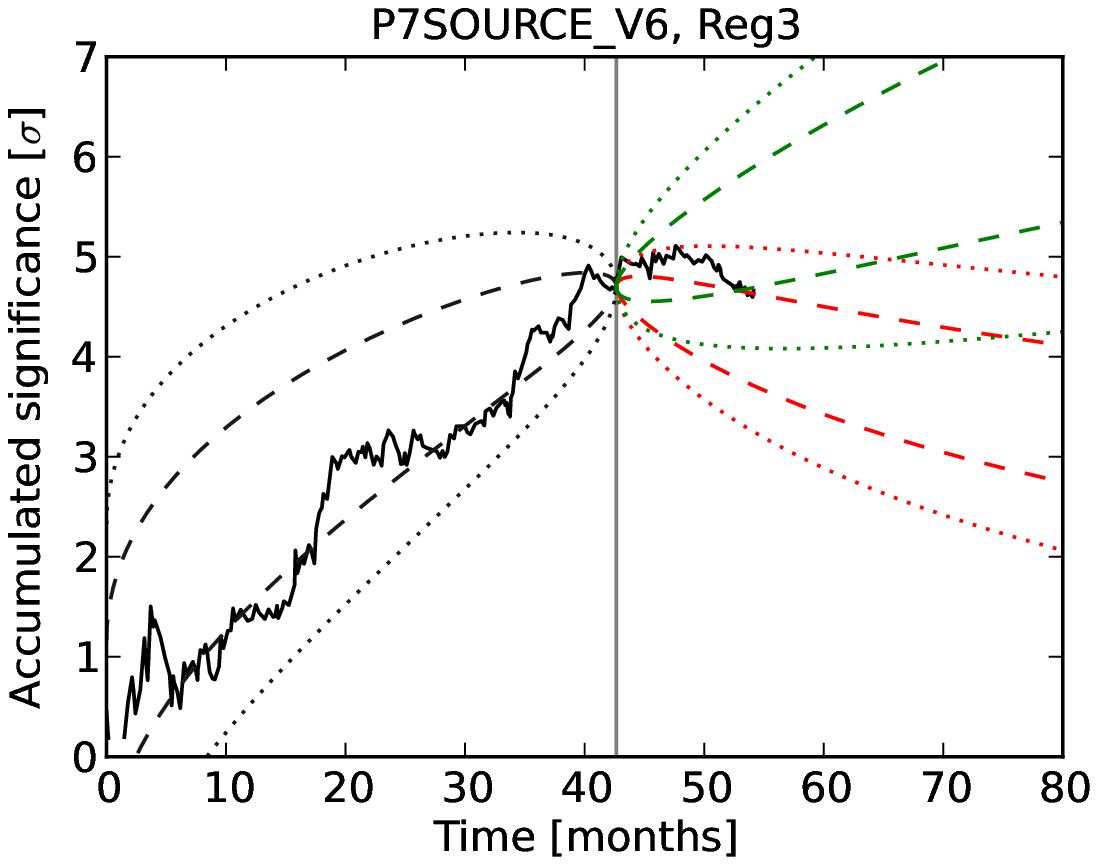}
  \includegraphics[width=.4\textwidth]{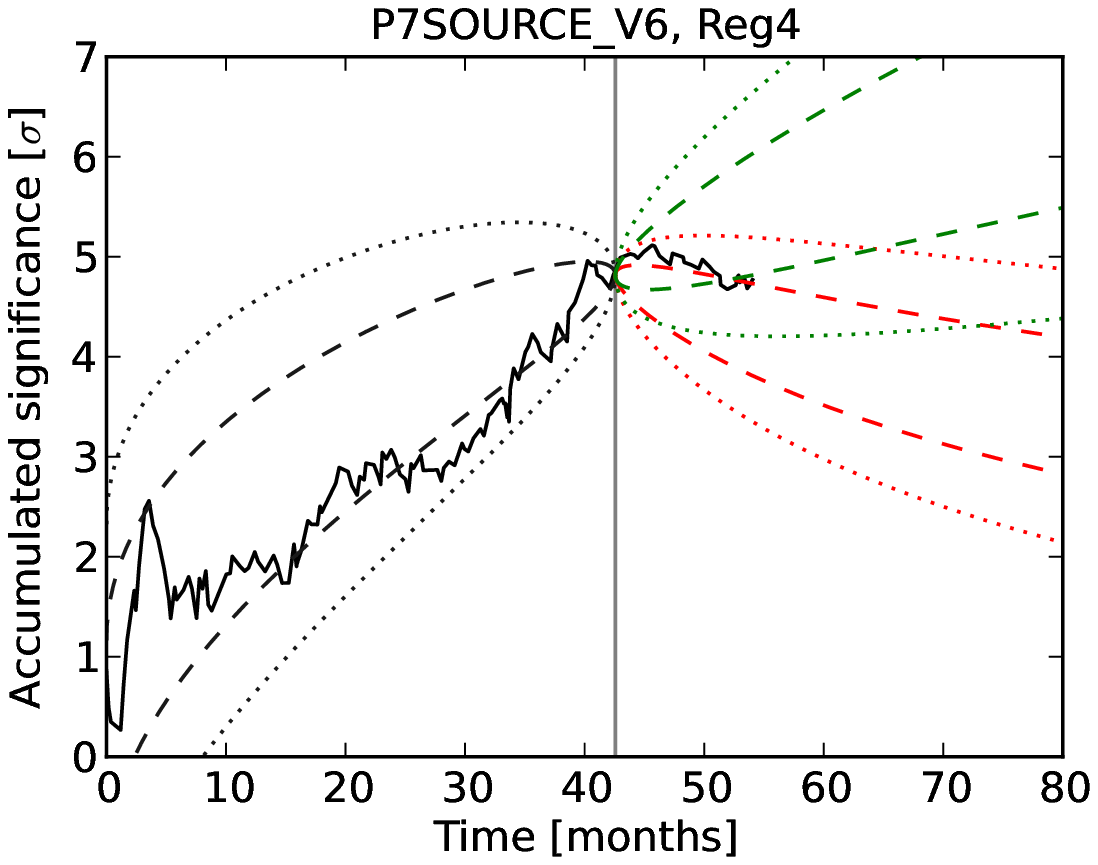}
  \caption{Time evolution of the accumulated significance of the line feature
  in Gaussian sigma in comparison with the expectations. The dashed (dotted)
  lines show the $68\%$CL ($95\%$CL) bands corresponding to a real signal
  (green), a statistical fluke (red) and a steady source in the past (black).
  The solid line shows the actual behaviour of the feature in the LAT data. The
  vertical line indicates the $8^{\rm th}$ of March 2012, which we use as
  reference point for a new trial-free measurement. In the fit, the line energy
  is fixed to $E_\gamma = 129.8$~GeV (see text for details). The four panels
  show results for the ROIs Reg3 (\textit{left}) and Reg4 (\textit{right}) from
  \cite{Weniger:2012tx}, for \texttt{P7CLEAN\_V6} (\textit{top}) and
  \texttt{P7SOURCE\_V6} (\textit{bottom}) class events. Data until the 22$^{\rm
  nd}$~of February 2013 is taken into account.}
  \label{fig:TStime}
\end{figure*}

\section{Updates}
\label{sec:update}
Given the best-fit values for the gamma-ray line flux as determined from data
taken until the $8^{\rm th}$ of March 2012 (as in \cite{Weniger:2012tx}), one
can easily project how the signal significance should evolve as more data is
added. On average, since we are in the background limited regime, the
accumulated significance in units of Gaussian sigma should grow like $x(t)=
x_0\sqrt{t/t_0}$ in case of a true signal, and fall like $x(t) =
x_0\sqrt{t_0/t}$ in case of a statistical fluke; here $x_0$ is the significance
measured at time $t_0$.  Using the Gaussian approximation, the $68\%$CL error
bands around this mean trend can be estimated analytically in case a signal is
present and read
\begin{align} 
  x_{\rm real}(t) =&\nonumber x_0\sqrt{\frac{t}{t_0}}
  \pm\left[\left(1+\frac{S}{B}\right)\!\!\left(1-\frac{t_0}{t}\right)
  \vphantom{\left(\left(\sqrt{\frac{t}{t_0}}\right)^2\Delta\alpha^2\right)}
  \right.\\
  &+x_0^2\left.\left(\sqrt{\frac{t}{t_0}}-\sqrt{\frac{t_0}{t}}\right)^2
  \left(\frac{\Delta S}{S}\right)^2\right]^\frac12.
  \label{eqn:nobelEvolution}
\end{align} 
In the present case, the signal-to-background ratio is $S/B\simeq35\%$, and
$\Delta S/S\simeq25\%$ is the statistical uncertainty of the measured line
flux.  If the signature is a statistical fluke, the expression reads instead
just
\begin{align} 
  x_{\rm fluke}(t) = x_0\sqrt{\frac{t_0}{t}}
  \pm\sqrt{1-\frac{t_0}{t}}\;.
  \label{eqn:fakeEvolution}
\end{align} 

In Fig.~\ref{fig:TStime}, we show how the accumulated significance of the
signature evolved over time (black solid line). In the fit, we fixed the
gamma-ray line energy to $E_\gamma=129.8$ GeV, and the fits are performed in an
energy range 65--260~GeV, which is slightly larger than in previous studies and
gives higher statistical power (we obtain similar results for
e.g.~40--400~GeV). As cut in \texttt{gtmktime} we take \texttt{DATA\_QUAL==1}
like in \cite{Weniger:2012tx}, but checked that using the recommended
\texttt{DATA\_QUAL==1 \&\& LAT\_CONFIG==1 \&\& ABS(ROCK\_ANGLE)<52} instead
does not significantly affect our results. The vertical bar indicates the
8$^{\rm th}$ of March 2012, which we use here as a starting point for a new
trial-free measurement. We show results for the ROIs Reg3 and Reg4 which gave
the highest significances in \cite{Weniger:2012tx}, and for
\texttt{P7CLEAN\_V6} and \texttt{P7SOURCE\_V6} event classes separately. The
dashed (dotted) lines indicate the $68\%$CL ($95\%$CL) bands that correspond to
the expected behaviour of a real effect (green) and a fluke (red); the
corresponding black lines show the expectations for a steady source in the
past. Green and red bands are respectively derived from the above
Eqs.~\eqref{eqn:nobelEvolution} and~\eqref{eqn:fakeEvolution}, the black band
follows from similar considerations. 

In case of \texttt{CLEAN} events, the curves are still compatible with a true
signal at the $95\%$CL, although the trend is clearly more pointing towards a
statistical fluke.  For \texttt{SOURCE} class events, the situation is similar,
although here the trend is less pronounced and the current significance lies
exactly between the expectations for a fluke and a signal.  We caution not to
overinterpret these figures: A decrease of the significance is possible over a
short period of time even for a real signal. But, if the significance continues
to drop at this rate, 6--12 month of additional data should be enough to
disqualify the 130 GeV feature from being a steady monochromatic line on top of
a power-law background.

We emphasize that following the time evolution of the significance within the
ROIs that were used in the initial analysis (and which were well defined
\emph{a-priori} regions optimized for dark matter searches, without any
optimization on the target sample itself) is the cleanest way to access the
statistical significance and stability of the signature. It allows a clear
prediction for the time evolution of the tentative line signal, free of any
hidden trials, and free of ambiguities in choosing ROIs and details of the
fitting methods. And with the accumulation of more data, the signature will
either pass that test or it will fail.

\section{Conclusions}
\label{sec:conclusions}
Our recent identification of a line-like signature around 130 GeV in the
Galactic center data of the Fermi-LAT raised an enormous interest, as it could
be the long awaited smoking gun signature for annihilation of WIMP dark matter
particles in the Universe.  The signature was independently confirmed by many
groups, and numerous studies accessed model-building aspects of the tentative
line signal, studied possible instrumental indications, and searched for
corroborating evidence from Galaxy clusters and dark matter subhalos.

Since the initial papers, exactly one year has passed by now, and it is time to
discuss how the signature evolved during that year. For the ROIs that we used
in the early papers, the expected signal event rate is about $\rm\sim1/month$,
and one finds clear, trial free and unambiguous predictions for how the
signature should behave if it is a real monochromatic line on top of a
power-law background. We confront these predictions with the data in
Fig.~\ref{fig:TStime}. In case of \texttt{P7CLEAN\_V6} events, the time
evolution of the accumulated significance clearly points more towards a
statistical fluke; in case of \texttt{P7SOURCE\_V6} events, it lies exactly
between the expectations for a real signal and a fluke. In all cases, the time
evolution is still compatible with a real signal at the $95\%$CL. However,
should the significance continue to drop at the same rate, 6--12 month of
additional data will be enough to exclude the possibility of a real gamma-ray
line signal with high confidence.

The release of reprocessed \text{P7} data by the Fermi-LAT team is expected to
happen soon, and it will allow a fresh look at all aspects of the 130 GeV
feature. The Air Cherenkov Telescope HESS-II should be able to confirm or rule
out the signature with high significance with less than 100 hours of Galactic
center observations (\cite{Bergstrom:2012vd}), but results are unlikely to be
released before late 2014. The availability of \texttt{P8} events, based on a
set of completely rewritten event reconstruction algorithms for the LAT, is
anticipated for later this year. It will likely become a landmark for deciding
whether one should give up on the 130 GeV feature, or whether it becomes
imperative to further investigate it with dedicated observations of the
Galactic center in the upcoming years.

\begin{acknowledgments}
  The author thanks the organizers of the \emph{4$^{\it  th}$Fermi Symposium},
  Monterey, California, 28 Oct -- 2 Nov 2012, for a stimulating conference, and
  Torsten Bringmann, Doug Finkbeiner and Meng Su for fruitful collaboration on
  this subject.
\end{acknowledgments}

\bibliography{/home/weniger/Dropbox/Archiv/archiv.bib}
\bibliographystyle{apsrev}

\end{document}